\documentclass[54pt]{article} 

\usepackage{geometry}
\usepackage{hyperref}
\hypersetup{
    colorlinks=true,
    linkcolor=blue,
    filecolor=magenta,      
    urlcolor=cyan,
    pdftitle={Overleaf Example},
    pdfpagemode=FullScreen,
    }
\usepackage[sort&compress,numbers]{natbib}

\title{A fermion phenomenology of low-temperature strongly-noncrystalline solids}

\author{Mihail Turlakov  (mihail.turlakov@gmail.com) }
\date{\today}

\begin{document}
\begin{large}
\maketitle

\section*{Abstract}

Insisting on the relevance of spin-statistics theorem, I propose that anomalous low-energy excitations (LEEs) of strongly-noncrystalline solids (SNSs), observed at low temperatures $T < 1 K$, are \textit{fermions, which are localized and weakly interacting.} This phenomenological theory rationalizes all low-energy quantum many-body states of SNSs as Goldstone bosons-phonons and half-integer-spin fermions. 
\textit{Fermi glass of LEEs, rather than statistical ensemble of isolated two-level systems, appears to be a consistent theory} of various thermal and nonlinear response properties of SNSs. 
A robust consequence of the theory is nearly constant-in-energy density of fermion states near Fermi energy, which in turn implies linear-in-temperature specific heat and dimensionless frequency-independent sound absorption. 
Consistent parameters of Fermi glass are estimated from theoretical scaling arguments as well as directly from experimental data.

\tableofcontents

\section{Introduction} \label{intro}

Although a lot of theoretical and experimental research on glasses and strongly noncrystalline solids has been done from the early 1970s, robust understanding of their low-temperature universal properties at temperatures $T < 1~K$ is missing. 
There are several paradoxes at the heart of the widely-adopted theoretical picture of the low-temperature properties of strongly noncrystalline solids (SNSs). This as-of-yet phenomenological picture of Tunneling Two-Level States (T-TLSs), proposed by Anderson-Halperin-Varma and Phillips (AHVP)\cite{anderson, Phillips}, assumed arbitrary low-energy states, which have a local tunneling nature and \textit{yet sufficiently large density of states}. These two-level states (TLSs) are thought paradoxically to remain local T-TLSs, although a basic estimate gives large interaction energy $E_{int}  \sim 10~K$, greater than typical TLS energy $E_{TLS}$,  between local T-TLSs at typical distances $\xi \sim 30 \dot{A}$ between tunneling units\cite{AJL-universality}. 
The universality of the dimensionless sound attenuation in a whole group of these different materials\cite{Pohl-review} below the temperature $T_Q \simeq 1~K$ accentuates the deep puzzles of low-temperature SNSs. In particular, the relation $l_s \approx 150 \lambda_s$ between mean-free path $l_s$ and the wavelength $\lambda_s$ of the phonons\cite{Freeman-Anderson} is almost independent (i.e. universal) from the detailed chemical and structural nature of SNSs. 

Strictly local picture of T-TLSs is problematic for several more reasons in addition to the neglect of TLS interactions mentioned above. First, two neighbouring T-TLSs have overlapping quantum wave-functions, and this consideration is neglected in AHVP phenomenology. Second, long relaxation times of AHVP theory are due to the long local-tunneling times rather than due to collective relaxation mechanisms associated with long length-scales and typical of glassy frustrated systems. Third, AHVP theory postulates many parameters - the magnitude and nearly constant-in-energy dependence of the energy density of T-TLSs as well as the distribution of relaxation times - in order to fit the universality of the sound attenuation\cite{Pohl-review}.
To put the third reason differently, AHVP model has too many free fitting parameters than a respectable theory should have in order to explain the remarkable universality.


I will argue in detail below that low-energy excitations (LEEs), which are responsible for linear-in-temperature specific heat for $T < 1 K$ experimentally, must be described as fermions. These composite fermions are localized due to quenched disorder and are an alternative basic ansatz to T-TLSs or TLSs of AHVP phenomenology. 
At long wavelengths (significantly longer than typical inter-atomic distances $a$), familiar transverse and longitudinal phonons are excitations-bosons of any rigid matrix of atoms. Similarly, for any disordered network of atoms but not under fully understood constraints, fermions appear as ubiquitously encountered LEEs in SNSs at low temperatures. 

In this paper I insist that the quantum mechanical overlaps between localized LEEs are non-negligible and determine the physical picture qualitatively. 
These localized LEEs - fermion quasi-particles - can be called glassons from the contraction of two words, glass and fermions. 
I also use the term of \textit{strongly} non-crystalline solids (SNSs) intentionally throughout this paper. Namely, the density of LEEs in SNSs must be sufficiently large, and therefore quantum-mechanical overlaps and exchanges between LEEs are non-negligible. 
Indeed, mutual quantum statistics of LEEs matters only at sufficiently large density.
SNSs include various classes of materials (glasses, amorphous solids, strongly irradiated crystals, etc.), where the density of LEEs is dense in a certain sense.

The nearly constant-in-energy density of LEEs, observed in the specific heat\cite{Zeller-Pohl} and in the internal sound friction\cite{Pohl-review}, is an important experimental clue, and the fermion phenomenological theory explains this fact more robustly and straightforwardly than AHVP theory. Namely, since fermions fill in fully ``Fermi sea" of all energy states below Fermi energy, the density of fermion states in the energy space just above finite Fermi energy, due to Fermi surface manifold degeneracy in the momentum space known for Fermi liquids, is naturally nearly constant as a function of energy.

In the section (\ref{Hamiltonian}), I write down explicitly an effective many-body low-energy Hamiltonian and discuss the relevant energy scales and coupling constants.
Theoretical scaling arguments of the section (\ref{length-scaling-crossover}) show that weak fermion-fermion and fermion-phonon interaction are consistent with the scaling of phonon, fermion, and fermion-fermion dipole energies as a function of length scale.

In the part (\ref{basic-consequences}), I follow up with several calculations of specific heat, thermal conductivity as well as the corrections to the glassons' density of states due to their interactions. The part (\ref{sound-attenuation}) summarizes the calculation of the internal sound friction $Q^{-1}$ and the reasons for its universality.
While the calculations in (\ref{basic-consequences}) are straightforward, the calculation in (\ref{sound-attenuation}) highlights the fact why the ratio $\lambda_s/l_s$ is frequency and temperature independent very naturally in the Fermi glass theory of SNSs.


The parameters of localized Fermi liquid, so called Fermi glass\cite{Freedman-Hertz}, are calculated  directly from the experimental data in the part (\ref{Fermi-parameters-experiments}). 
This basic calculation highlights that, by guessing a plausible Fermi energy $\epsilon_F \simeq 10~K$ and taking specific heat slope $C_v/T$ from experiments, glasson Fermi parameters are reasonable without any further fitting parameters. Indeed, the same glasson parameters are also calculated from theoretical scaling arguments in the part (\ref{length-scaling-crossover}).
The question "why fermions, not TLSs?" is discussed in the section (\ref{fermions-not-TLS}), and the answers to these questions summarize the most essential and novel elements of the fermion phenomenology of SNSs.
In the part (\ref{open-questions}), I discuss not only open questions beyond fermion phenomenology but also the motivation for further experimental probes. 


\section{Glasson phenomenology and nearly constant-in-energy density of states} \label{phenomenology}

I start by motivating the Fermi-statistics nature of LEEs. 
At zero temperature, spin-statistics theorem tells us that only bosons and fermions can exist. I reckon that LEEs can only be localized fermions since 
localized bosons would hybridize \textit{strongly} with phonons, extended boson states. Extended state of a phonon can coexist with a localized state of glasson because they have different quantum numbers of spin. Phonons have zero spin, while glassons-LEEs are spin-half quasi-particles. Moreover, weak coupling between glassons and phonons can be expected on the symmetry grounds, because they are different species which couple (hybridize) only weakly with each other.

In the phenomenological style of the theories of Fermi liquid of Helium-3 and Bose liquid of Helium-4 by Lev Landau, I propose that all LEEs in SNSs are two weakly interacting particle species, propagating bosons-phonons and localized fermions-glassons. An important difference from the aforementioned theories, in particular, from liquid Helium-3, is that glasson is a composite fermion, which possibly does not have a well-recognised and elementary high-energy fermion ``parent'' (yet see the last paragraph of the section (\ref{open-questions}) for a possible ``parent'' interpretation). For comparison, Landau-Fermi quasi-particles in Helium-3 do have ``parents'' as $He^3$ atoms, yet phonons have never had elementary ``high-energy parents''. 

Low-energy Hamiltonian is postulated in the part (\ref{Hamiltonian}) with plausible parameters. The ground state of this Hamiltonian is Fermi glass\cite{Freedman-Hertz}, which is the ground state of 
Anderson model of localized fermions in the context of SNSs at zero temperature. 
By increasing the disorder strength from the limit of weak or no disorder to the limit of strong disorder, normal Landau-Fermi liquid can be `adiabatically transformed' to Fermi glass. This `adiabatic transformation' keeps the same constant-in-energy density of fermion states near the Fermi energy $\epsilon_F$. 
Theoretical scaling arguments as a function of the length scale - the part (\ref{length-scaling-crossover}) - highlight weak-coupling Fermi glass picture of SNSs as well as the consistency of the same aforementioned single crossover length scale $\xi$.

Non-negligible quantum overlaps between the same type of (even though localised) excitations imply the necessity of definitive quantum statistics, due to the spin-statistics theorem. Virtual exchanges of indistinguishable particles in the perturbation theory must respect either boson or fermion commutation relation. Since fermions are localised, Fermi statistics is not expressed explicitly in the wave-function symmetry under permutations in the real space, but rather in the virtual exchanges calculated in the perturbation theory.

Two states of an isolated TLS are equivalent to unoccupied and occupied states of a single fermion. The same fact can be expressed mathematically because there is exact mapping between fermion creation and annihilation operators and Pauli matrices describing a TLS. 
In this particular sense, in the introduction I stated that fermions replace TLSs in the context of many-body quantum states.
However it is more important how to describe an ensemble of non-isolated TLSs.
Fermions can be thought as a particular version of an ensemble of local LEEs, when there are non-negligible quantum-mechanical overlaps between the localized states.

\subsection{Low-energy Hamiltonian and energy scales of SNS} \label{Hamiltonian}

There is a clear hierarchy of energy scales in SNSs. The highest energy scale $\Lambda \sim 10~ eV \sim 10^5~K$ (or rather a range of $1-10~ eV$) is the chemical inter-molecular and inter-atomic bonding which is also the local interaction strength. 
Debye energy $E_D \sim 10^3~K$ is the next energy scale lower than the chemical bonding energy. Debye energy determines the upper energy band of any local and phonon-like excitations.

Medium-range order\cite{Elliott} (MRO) length $\xi \sim 30~\dot{A}$ defines the energy scale $E_{int} \sim 10~K$ by the relation $E_{int} \simeq \hbar c_t/\xi$, where $c_t$ is the transverse sound velocity (for instance, in vitreous silica $c_t \simeq 4 \cdot 10^5 cm/s$). The elastic dipole coupling energy of local modes $\frac{\Lambda^2}{\rho c_t^2} (1/\xi^3)$, separated by MRO distance $\xi$, is also about the energy $E_{int}$ from the relation $\frac{\Lambda^2}{\rho c_t^2} (1/\xi^3) \sim (\hbar c_t)/\xi$\cite{Yu-Leggett}. 

The lowest collective energy scale $E_Q$ is given by the quantum crossover temperature observed as a crossover from linear-in-temperature at $T < 1 K$ and cubic-in-temperature at $T > 10 K$ dependencies of the specific heat as well as the plateau in the thermal conductivity. To be specific, I define this quantum (therefore subscript $Q$ in $E_Q$) energy scale $E_Q=k_B T_Q=1~K$, which is about a factor of $10^3$ lower than Debye energy $E_D$. 


Effective low-energy Hamiltonian has to describe a ground state and all low-energy excitations with the transition matrix elements between these different excitations. 
The full Hamiltonian is the sum
\begin{equation}
\hat{H}_{total} = \hat{H}_{main} + \hat{H}_{int},
\label{H_total}
\end{equation}
where $\hat{H}_{main}$ and $\hat{H}_{int}$ are the main and interaction parts of the Hamiltonian correspondingly.
Quantum many-body states are better expressed in the second quantization notations. 
The main part $\hat{H}_{main}$ of the Hamiltonian describes glassons and phonons

\begin{equation}
\hat{H}_{main}=\sum_{\alpha}
\epsilon_{\alpha}\,\hat{c}^{\dagger}_{\alpha}\hat{c}_{\alpha}+
\sum_{\alpha \neq \beta}
t_{\alpha,\beta}\,\hat{c}^{\dagger}_{\alpha}\hat{c}_{\beta}+
\sum_{k}
\omega_{k} \hat{b}^{\dagger}_{k}\hat{b}_{k}
+
\sum_{k}
A_{k, k'} \hat{b}^{\dagger}_{k}\hat{b}_{k'}.
\label{H_main=}
\end{equation}
The creation and annihilation operators of glassons are  $\hat{c}^{\dagger}_{\alpha}$ and $\hat{c}_{\alpha}$, while
the creation and annihilation operators of phonons are
$\hat{b}^{\dagger}_{k}$ and $\hat{b}_{k}$.
Glassons and phonons are well-defined energy resonances of the same solid medium. 
The interactions are weak if quasiparticle energies are small $(\epsilon, \omega) < E_{int}$.
Let me call for brevity the proposed here theory as a glasson-phonon phenomenology (GPP). 

The fermion part of non-interacting $\hat{H}_{main}$ is simply Anderson model studied widely in the research of single-particle localization.
It is natural to think that indices $\alpha$ and $\beta$ describe states in the real space, while $k$ and $k'$ are the labels in the momentum state.
The energies $\omega_k$ describe both, longitudinal $\omega_k = c_l k$ and transverse $\omega_k = c_t k$, phonons.
The local fermion energies $\epsilon_\alpha$ are distributed in a certain wide range $W \gg E_Q$ due to locally different structure of the blobs. 
We assume that the disorder is strong for glassons-fermions, namely that hopping is weak $t \ll W$. Reasonable, but quite uncertain, estimates are the bandwidth $W \simeq 100~K$ and the hopping parameter $t \leq 0.1-1~K$.
In reality, both quantities $\epsilon_\alpha$ and $t_{\alpha,\beta}$ are likely to be widely distributed (disordered) for SNSs. The disorder-induced matrix element $A_{k, k'}$, which describe elastic scattering of phonons from disorder, becomes non-vanishing only when $(k, k') \geq \xi^{-1}$. 

The interaction part $\hat{H}_{int}$ of the Hamiltonian describes glasson-glasson, glasson-phonon, and multi-phonon anharmonic interactions. Let us write down explicitly glasson-phonon interaction of two types

\begin{equation}
\hat{H}_{int}^{gl-ph}=  \sum_{\alpha, \beta , k}
\sqrt{\omega_k} M_{\alpha\beta, k}^z 
\hat{c}_{\alpha}^{\dagger}\hat{c}_{\beta}
(\hat{b}^{\dagger}_{-k} + \hat{b}_{k})
+
 \sum_{\alpha , k} \sqrt{\omega_k} M_{\alpha, k}^x 
(\hat{c}_{\alpha}^{\dagger}+ \hat{c}_{\alpha})
(\hat{b}^{\dagger}_{-k} + \hat{b}_{k}),
\label{H_int}
\end{equation}
where $M_{\alpha\beta, k}^z$ and $M_{\alpha, k}^x$ are constants in the long-wave limit $k \rightarrow 0$. The second term containing $M_{\alpha, k}^x$ is allowed due to the lack of fermion-charge conservation. These interactions are familiar coupling terms expressed via fermion operators rather than Pauli matrices $\sigma^z$ and $\sigma^x$ for TLSs. 
\textit{The main conjecture of the paper} expressed mathematically is that for different indices $\alpha \neq \beta$ anti-commuting relation $c_\alpha c_\beta + c_\beta c_\alpha =0$ holds rather than commuting relation assumed in the Standard Model\cite{anderson, Phillips}. AHVP proposed that local LEEs, being equivalent to pseudo-spins, obey bosonic commutation between any local field operators, describing different local sites, because TLSs are assumed to be independent.

\subsection{Single crossover length and weak coupling} \label{length-scaling-crossover}

Low-energy Hamiltonian, Eqn.(\ref{H_total} - \ref{H_main=}), implies the ground state of Fermi glass.
Yet the important low-energy parameter, which is determined by high-energy physics and not explicit in the Hamiltonian, is the glassons density in the ground state. In what follows I estimate the density of glassons, or equivalently Fermi volume of glassons, by identifying a single crossover length with the Fermi wavelength of glassons. This crossover length is determined by matching principal involved energies. The crossover length $\xi$ marks the onset of long-distance universality for $R > \xi$. The same considerations illustrate the onset of weak coupling of fermion-fermion and fermion-boson interactions. 

Initially it is insightful to compare how different energies, involved in the problem, scale as a function of length scale $R$ of an amorphous solid blob. 
We have the phonon energy $E_s = \frac{\hbar c_t}{R}$, the kinetic localization energy (of glasson) $E_{kin} = \frac{\hbar^2}{2 m_{gl} R^2}$, and the dipole interaction energy $E_{int-dipole} = \frac{\Lambda^2}{\rho c_t^2 R^3}$ between two blobs (blocks).
For sufficiently large length scales $R \gg \xi$, we have necessarily the following inequalities

\begin{equation} \label{inequalities}
    \frac{\hbar c_t}{R} \gg \frac{\hbar^2}{2 m_{gl} R^2} \gg \frac{\Lambda^2}{\rho c_t^2 R^3}.
\end{equation}
The inequality $E_s \gg E_{int-dipole}$ implies weak anharmonic phonon-phonon coupling due to a phonon energy being larger than dipole interaction between two blobs, if excited by two different phonons. 
The inequality $E_s \gg E_{kin}$ implies weak phonon-fermion coupling due to a energy-momentum mismatch between a phonon and a localized fermion.

The inequality $E_{kin} \gg E_{int-dipole}$ suggests weak fermion-fermion coupling due to a local fermion excitation energy being larger than dipole interaction between two fermions-blobs. To put it differently, a localised fermion is a well-defined local resonance (i.e. excitation).
In particular, the ratio of the glasson kinetic energy and glasson interaction energy is
\begin{equation}
    \frac{E_{kin}}{E_{int-dipole}} = \frac{R}{a_{gl}} = \frac{1}{a_{gl} n_{gl}^{1/3}},
\end{equation}
where $a_{gl} = 2 \Lambda^2 m_{gl}/(\hbar^2 \rho c_t^2)$ and the glasson density $n_{gl} \simeq 1/R^3$ for inter-particle distance $R$. It is clear that at sufficiently dilute densities $a_{gl} n_{gl}^{1/3} \ll 1$, glassons, interacting via dipole interaction, are in the regime of the weak interaction since $E_{kin}/E_{int-dipole} \gg 1$. This situation is opposite to the situation of electrons interacting via Coulomb interaction at dilute densities in three dimensions, where  $E_{kin}/E_{int-Coulomb} = a_B/R = a_B n_e^{1/3} \ll 1$ ($a_B$ is Bohr radius) is the strong interaction limit for $n_e \rightarrow 0$. 

Now let us make a simple, but a far reaching, \textit{assumption that there is a single crossover length scale $\xi$} so that 

\begin{equation} \label{equality1}
    \frac{\hbar c_t}{\xi} \simeq \frac{\hbar^2}{2 m_{gl} \xi^2},
\end{equation}
\begin{equation} \label{equality2}
    \frac{\hbar c_t}{\xi} \simeq \frac{\Lambda^2}{\rho c_t^2 \xi^3}.
\end{equation}
Namely, that all three energies - $E_s$, $E_{kin}$, and $E_{int-dipole}$ - become roughly the same at the single length scale $\xi$. For $R \gg \xi$, these energy scales become clearly hierarchically ordered as $E_s \gg E_{kin} \gg E_{int-dipole}$.
In the introduction of the part (\ref{Hamiltonian}), we already noticed that MRO $\xi$ is consistent with the scaling relation of Eqn.\ref{equality2}. Appealing to additional scaling relation of Eqn.\ref{equality1}, we can estimate the parameter $m_{gl}$, which has the dimensionality of physical mass.

Defining a constant $H \equiv \sqrt{\hbar \rho c_t}$, which is dependent \textit{only} on long-wavelength properties $\rho$ and $c_t$, Eqn's (\ref{equality1}, \ref{equality2}) give the following simple expressions for the length $\xi$ and the mass $m_{gl}$

\begin{equation} \label{xi-m}
    \xi = \frac{\Lambda}{c_t H}, ~~~~ m_{gl}= \frac{\hbar H}{2 \Lambda}.
\end{equation}
Numerical estimates for vitreous silica give $H \simeq 0.9*10^{-10} g/s$, $\xi \simeq 44 \dot{A}$, and $m_{gl} \simeq 5.6*10^{-26} g$. To be concrete, I used $\Lambda \simeq 10eV$ while more realistic values are only several $eV's$.
The crossover energy is $E_\xi \sim 10K$ (order of magnitude).

The theoretical expressions - Eqn's (\ref{xi-m}) - for the length $\xi$ and the mass $m_{gl}$ are simple and general. The qualitative interpretation seems that, 
\textit{under some conditions, disorder of chemical bonding can conspire to medium-range order length $\xi$ and mass parameter $m_{gl}$.} 
These expressions depend only on four parameters, two experimentally measured macroscopic parameters $\rho$ and $c_t$, chemical bonding energy $\Lambda$, and fundamental Planck's constant $\hbar$. 

Notice that so far in this section the quantum localisation of a particle was assumed, but not a particular quantum statistics. 
In the section \ref{Fermi-parameters-experiments}, I will further assume that localised particles have Fermi statistics.

The simplicity of  Eqn's (\ref{xi-m}) may appeal to the ubiquity and universality of LEEs in various disordered solids, but the limitations must be stated. Sufficient and necessary conditions for the existence of dense LEEs are still not known very well structurally and chemically.
Experimentally not all disordered solids posses sufficiently large density of LEEs, and it is important to research further into the microscopic models and constraints on the applicability of Eqn's (\ref{xi-m}). For instance, some large variation of local density, chemical bonding strengths, and geometry of random-network chemical bonds, not explicit in Eqn's (\ref{xi-m}), seem to be required\cite{Phillips-review}. 

In the section (\ref{Fermi-parameters-experiments}), I calculate the mass $m_{gl}$ and Fermi wave-vector $k_F$ essentially from the experimental data.
The consistency of the results for $m_{gl}$ and $k_F$ within a factor of 2-3, derived from experimental data and from theoretical scaling calculations in this part of the paper,  is definitely encouraging and likely not accidental. Notice that there are only two `free parameters' $\epsilon_F$ and $\Lambda$ in theoretical and experimental arguments taken together, but even these two parameters can be reasonably estimated and can vary only within one order of magnitude at most.

\section{Basic results of the phenomenology} \label{basic-consistency}

In this part of the paper, I calculate the basic consequences of the glasson-phonon phenomenology (GPP). 


\subsection{Basic results for the specific heat and thermal conductivity} \label{basic-consequences}

At the basic level of GPP, core experimental results of SNSs at $T < T_Q$ can be easily understood. 
Some basic calculations, which follow, are analogous to the calculations of AHVP model\cite{anderson, Jackle}, since a localised fermion-glasson can be mapped exactly into a TLS via the mapping from fermion operators to Pauli matrices\cite{JW-transformation}. 

Firstly, localized glassons for $T \ll \epsilon_F$ will exhibit linear-in-temperature specific heat as any localized Fermi glass and normal Landau-Fermi liquid\cite{Freedman-Hertz} does. This contribution is as usual due to the thermally excited number $n_F (T) \sim T/\epsilon_F$ of fermions, and this fermion contribution can exceed phonon contribution to the specific heat being proportional to $T^3$. 

Secondly, the phonon thermal conductivity is  $\kappa_{ph} (T) \sim C_{ph} (T) \tau_{k, ph} (T) \sim T^2$, due to phonon-glasson scattering. The thermal conductivity was shown experimentally to be completely dominated by phonons\cite{Zaitlin-Anderson} for $T < T_Q$ and to have roughly square-in-temperature dependence. From the kinetic theory\cite{Abrikosov-book} we have for the thermal conductivity of phonons

\begin{equation} \label{kappa-thermal}
    \kappa_{ph} = \frac{1}{3} C_{ph} (T) c_s^2 \tau_{k, ph},
\end{equation}
where $C_{ph} \sim T^3$ is the phonon specific heat for $T \ll T_D$ below Debye energy $T_D$, $c_s$ is the sound velocity, and $\tau_{k, ph}$ is the momentum-scattering time of phonons. The inverse scattering time $\tau_{k, ph}^{-1}$ is proportional to $Q_{eff} n_F(T)$, where the quantity $Q_{eff}$ is the effective glasson-phonon scattering cross-section. 
Glassons are point scatterers of phonons, since de Broiglie wavelength of phonons is much larger $\lambda_s \gg \lambda_F$ than the analogous quantity for glassons $\lambda_F$. The number of point scatters $n_F (T)$ is proportional to the temperature $T$. Therefore, qualitatively we do get  the quadratic temperature dependence of the phonon thermal conductivity $\kappa_{ph} (T) \sim C_{ph} (T) \tau_{k, ph} (T) \sim T^2 $ in the leading order.

Glassons themselves are localised and can contribute to the thermal conductivity only through thermally activated hopping. We do not calculate this subleading contribution to the thermal conductivity here.

Quantitatively, we can calculate the inverse scattering time $\tau_{k, ph}^{-1}$ from Fermi Golden rule explicitly for the interaction from Eqn. (\ref{H_int})

\begin{equation}
    \frac{1}{\tau_{k, ph}}= \frac{2\pi}{\hbar} \vert H_{int}^{gl-ph} \vert^2 \delta(E_f - E_i),
\end{equation}
where $E_i$ and $E_f$ are initial and final phonon energies. 
The first term in Eqn. (\ref{H_int}) gives

\begin{equation}
   \frac{1}{\tau_{k, ph}^z}= \frac{2\pi \omega_k}{\hbar} \vert M_{\alpha\beta, k}^z \vert^2 \int d\epsilon_1 (1- n_F (\epsilon_1/T)) n_F(\epsilon_2/T) \sim T \omega_k 
\end{equation}
where two fermion energies are related by the energy conservation $\epsilon_1 - \epsilon_2= \hbar \omega_k$.
The second term in Eqn. (\ref{H_int}) gives

\begin{equation}
   \frac{1}{\tau_{k, ph}^x}= \frac{2\pi \omega_k}{\hbar} \vert M_{\alpha\beta, k}^x \vert^2 (1- 2 n_F (\omega_k/T)) \sim \omega_k tanh \left( \frac{\omega_k}{2T} \right)
\end{equation}
where $\epsilon=E-\epsilon_F = \hbar \omega_k$ is a fermion energy relative to the Fermi energy $\epsilon_F$. For the thermal phonons such that $\omega_k \sim T$, the dominant scattering mechanism, due to the term $M_{\alpha\beta, k}^x$, is $1/\tau_{k, ph}^x \sim T$. Thus, once again, the phonon thermal conductivity is $\kappa_{ph} (T) \sim C_{ph} (T) \tau_{k, ph}^x (T) \sim T^2 $.

Thirdly, I calculate the marginal corrections of weak glasson-glasson interactions to the energy density of states and consequently to the glasson specific heat and phonon thermal conductivity. 
Due to dipolar, weak-coupling but long-range, fermion-fermion interaction, falling as $\alpha_{d-d}/R^3$ with the distance $R$, the non-interacting energy density of states $\nu_0 (\epsilon) = p_F m_F/\hbar^3 = \rho C_v (T) /(k_B T) $ is suppressed logarithmically as a function of energy and temperature.
Such a `dipolar gap' suppression is due to the same mechanism as Efros-Shklovskii Coulomb gap mechanism\cite{Efros-Shklovskii}. The corrected density of states $\nu (\epsilon)$ is

\begin{equation}
    \nu (\epsilon) = \frac{\nu_0 (\epsilon)}{1 + \nu_0 (\epsilon) \alpha_{d-d} ln(E_Q/T)}.
\end{equation}

Therefore the temperature dependence of the specific heat becomes $ C_{v, gl} \sim T/ln(T_Q/T)$  under the condition $\nu_0  \alpha_{d-d} ln(E_Q/T) \gg 1$. This is again a very crude familiar argument, and more detailed analysis of the effect of angular dependence of dipole-dipole interaction is clearly desirable. This correction means that the specific heat can be approximately written as $\sim T^{1+\alpha}$ with small $\alpha \ll 1$, an experimental fit in some temperature range, instead of simple linear-in-temperature $\sim T$ dependence. Since the phonon mean free path is $l_s (T) \sim 1/n_F (T)$ due to the scattering of phonons by glassons of the number $n_F (T) \sim \nu (\epsilon) T$, then the corrected thermal conductivity becomes $\sim T^2 ln(T_Q/T) \approx T^{2 - \alpha}$ replacing $T^2$ dependence. These corrections are consistent with experiments\cite{Yu-Leggett}.

One of the most important and basic properties of glasses, and even glasses at low temperatures, is long energy relaxation times, for instance, revealed in the specific heat experiments\cite{Loponen-1982}. At the level of non-interacting Hamiltonian $\hat{H}_{main}$ (eqn. \ref{H_main=}), there is no energy relaxation. Since all types, glasson-phonon and glasson-glasson, of interactions $\hat{H}_{int}$ are weak and responsible for thermalization, the relaxation times can remain long. It is important to investigate in more detail how glasson-phonon and glasson-glasson induced relaxations are similar and different in Fermi glass versus TLS glass\cite{Imry-2017}.



\subsection{Universal sound attenuation at $T < 1~K$} \label{sound-attenuation}

The internal friction measurements\cite{Pohl-review} highlight clearly the universality of LEEs in SNSs. It is particularly intriguing to understand the internal friction coefficient $Q^{-1} (\omega)$, since this quantity is a direct measurement of small dimensionless interaction constant. The quantity $Q^{-1}= 1/(2 \pi) (\lambda/l) $ is expressesed by the ratio of phonon wavelength $\lambda (\omega)$ to its mean free path $l (\omega)$ for various frequencies $\omega$. In the regime $T < T_Q$ for SNSs, in a certain range of temperatures and frequencies, the internal friction is a small quasi-universal quantity of the order of $10^{-2}-10^{-3}$\cite{Pohl-review}. 

It is insightful to make a short excurse into the theory of sound attenuation by electron Landau-Fermi liquid in simple metals\cite{Abrikosov-book}. Similarly to SNSs, in simple metals the coefficient of sound internal friction is small and temperature and frequency independent. In particular, this coefficient is equal to $Q^{-1}_{LFL} \sim c_s/v_F \sim \sqrt{m_e/M_{ion}}$, where $c_s$ is the sound velocity and $v_F$ is a Fermi velocity for itinerant electrons. In simple metals, the internal friction is small and is of the order of $Q^{-1}_{metals} \sim 10^{-2}-10^{-3}$ due to the small ratio $c_s/v_F \ll 1$. In this case, the smallness of the dimensionless electron-phonon coupling, expressed by the internal friction $Q^{-1}_{metals}$ is ultimately due to the small ratio of the electron mass $m_e$ to the ion mass $M_{ion}$. It is also important to highlight that the result $Q^{-1}_{LFL} \sim c_s/v_F$ is equally valid independently from the parameter $\omega_s \tau_F = (c_s/v_F) (l_F/\lambda_s)$ being larger or smaller than 1\cite{Abrikosov-book}. 


Following through the kinetic equation calculations\cite{Abrikosov-book} or equivalently Fermi Golden rule under the condition $\omega_s \tau_{F, gl} \gg 1$, 
I get the expression for the dimensionless absorption 
\begin{equation} \label{Q-1}
    Q^{-1}_t  \simeq \pi \frac{\nu (\epsilon_F) \lambda_{gl-ph}^2}{\rho s^2} \frac{c_t}{v_F} \simeq \pi K_{F-s} \frac{c_t}{v_F},
\end{equation}
where the glasson-phonon coupling constant $\lambda_{gl-ph}^t= M^{x, t}_{\alpha \beta} \sqrt{c_t}$ from the definition in Eqn.(\ref{H_int}).
The dimensionless constant $K_{F-s}$ is given by the expression 
\[
K_{F-s} \equiv \frac{\nu (\epsilon_F) \lambda_{gl-ph}^2}{\rho c_t^2} = \frac{\lambda_{gl-ph}^2}{\epsilon_F M_{atom} c_t^2} (k_F a)^3.
\]
The question arises how the coupling constant $\lambda_{gl-ph}$ depends on various parameters of SNSs. I conjecture that $\lambda_{gl-ph}^2 \simeq \epsilon_F M_{atom} c_t^2$ scales similarly to the arguments used in simple metals\cite{Abrikosov-book}. Then the coupling constant  $\lambda_{gl-ph}$ for large glasson quasi-particles of size $R \geq \xi \gg a$ is significantly smaller than the bare coupling coupling constant $\Lambda$ for the local modes. Numerically, using the numbers $\epsilon_F=10~K$ and $M_{atom} c_t^2 \sim 10^5~K$, the coupling constant is $\lambda_{gl-ph} \simeq 10^3~K$. All estimates here are only crude order of magnitude. Thus $\lambda_{gl-ph} \sim 10^{-2} \Lambda$ is indeed much smaller than the bare chemical bonding energy $\Lambda$.

Finally the result for the internal friction coefficient of the transverse phonons in SNSs is
\begin{equation} \label{Q-transverse}
    Q^{-1}_t = K_{Q,t} (k_F a)^3 \frac{c_t}{v_F},
\end{equation}
where $K_{Q, t}$ is the number of order 1. This result corresponds to the resonant regime calculation of AHVP theory\cite{Jackle} in the limit $\omega/T \gg 1$. The internal friction for longitudinal phonons is similarly $Q^{-1}_l \sim (k_F a)^3$.
\textit{The ratio of the velocities $c_s/v_F$ in SNSs is very roughly of order 1.}
At the phenomenological level, small constant $Q^{-1}$ is due to small quantity $(k_F a)^3$, i.e. small dimensionless Fermi volume. 


\section{Glasson parameters and open questions} \label{consistency}

In the part (\ref{Fermi-parameters-experiments}), I present a basic calculation of glasson Fermi parameters directly from the linear-temperature slope of the specific heat and conjectured Fermi energy. 
In the part (\ref{fermions-not-TLS}) I summarize the arguments why universal LEEs are fermions rather than two-level states (TLSs). 
I conclude this section of the paper with a tentative discussion of  possible microscopic understanding of the nature of excitations-glassons and experimental tests (part \ref{open-questions}).

\subsection{Fermi parameters of glassons from experimental data} \label{Fermi-parameters-experiments}

As compared to the theoretical scaling arguments of the part (\ref{length-scaling-crossover}), we can deduce some parameters of Fermi glass (or Fermi liquid) essentially from experimental data directly. Since Fermi glass is `adiabatically connected' to Landau-Fermi liquid, I will use the standard relations from the Landau-Fermi liquid theory. 

The linear-in-temperature slope of the specific heat capacity is $(C_v/T) \sim p_F m_F$, while the Fermi energy is $\epsilon_F = p_F^2/(2 m_F)$. If we take from experiments the temperature slope of the specific heat $(C_v/T)$ and estimate the values of Fermi energy $\epsilon_F$, the above expressions make obvious that we can calculate separately the glasson Fermi mass $m_{gl}$ and the Fermi momentum $p_F=\hbar k_F$. 
Fermi wave-vector $k_F$, or equivalently de Broglie wavelength, has a meaning, that only the states at and below this wave-vector are occupied in the momentum space. 
Due to strong elastic scattering by disorder, glasson at the energy $\epsilon_F$ is localized in real space but it still has a well-defined energy, which is a linear superposition of multiple-direction waves with the wavevector $k_F$.  
\textit{`Fermi glass'} is simply the ensemble of non-interacting localized fermions\cite{Freedman-Hertz, Fleishman-Anderson} filling in the Fermi sea up to the Fermi energy $\epsilon_F$. Since the static strong disorder causes only elastic scattering, `adiabatic continuity' from normal no-disorder Landau-Fermi liquid to strongly-disordered Fermi glass as a function of disorder strength makes quasiparticle momentum not a good quantum number and yet keeps quasi-particle energy as a good quantum number. 

Linear-temperature dependence of the specific heat due the Fermi occupation number $n_F (T)$ becomes very clear when the temperature becomes smaller than (roughly) one tenth of Fermi energy. Since the linear temperature dependence of the specific heat in amorphous solids sets in below $1~K$, I estimate the Fermi energy to be $\epsilon_F = 10~K= 1.4 \cdot 10^{-15} erg$.
The experimental specific heat slope\cite{Zeller-Pohl} is $C_v/T \approx 10 ~erg/(g~ K^2)$.
We have the following theoretical expressions for the basic Landau-Fermi parameters

\begin{equation} \label{p_F}
k_F =  \left( 6 \epsilon_F \left( \rho \frac{C_v}{T} \right) \right)^{1/3},
\end{equation}

\begin{equation} \label{m_F}
m_{gl} = \hbar^2 \left( \frac{3}{2\epsilon_F} \left( \rho \frac{C_v}{T} \right)^2 \right)^{1/3}, 
\end{equation}
where $\rho$ is the mass density of a solid ($\rho \simeq 2.2 g/cm^3$ for vitreous silica). 
As an example, for vitreous silica, I estimate the numerical value of the Fermi wave-vector $k_F \approx 2.1*10^6 cm^{-1}$. For the reference, for localised glassons, Fermi parameters without direct physical meaning are the glasson mass $m_{gl} \approx 5*10^{-26} g \sim 10^{-2} m_{He^3}$ and Fermi velocity is $v_F \simeq 2.3*10^5 cm/s$.

Importantly, quantum de Broglie wavelength of glassons is $a_F \sim k_F^{-1} \sim 50 \dot{A}$, much longer than inter-atomic distances of $a \simeq 2.5 \dot{A}$ (for the sake of specific numerical value). 
Since $k_F a \ll 1$, the condition for coarse-graining of blobs of the size $a_F$ is satisfied. Namely, fermions-glassons in SNSs are a consistent concept. 

Since glassons are localized, the parameters $m_{gl}$ and Fermi velocity $v_F=p_F/m_{gl}$ do not have direct physical meaning. But Fermi volume and closely related Fermi momentum $p_F$ in the presence of only elastic scattering from disorder, should be measurable and do have physical meaning.

We can notice that glasson de Broglie wavelength $a_F$ is crudely equal to MRO length $\xi$, which is estimated as $\frac{\Lambda}{c_t H}$ in the section (\ref{length-scaling-crossover}). Namely, I guess \textit{an important relationship between Fermi wavevector $k_F$ and the crossover length $\xi$, which is $k_F \xi \sim 1$.}
The existence of a single crossover scale $\xi$, rather than two separate scales from inequalities (Eqn. \ref{inequalities}), might be one of the main criteria for the convergence to the universality class of SNSs.
Thus, \textit{LEEs are plausibly fermions, since glasson de Broglie wavelength from Eqn.(\ref{p_F}) is about the same as the crossover length $\xi$, derived from Eqn.(\ref{equality2}).} 


Using Eqns. \ref{xi-m}, we can also write a simple expression for the ratio of the Fermi energy $E_F=p_F^2/(2 m_F)$ and the crossover energy $E_\xi=\hbar c_t/\xi$ from Eqn.(\ref{equality2})

\begin{equation} \label{fermi-crossover-ratio}
    \frac{E_F}{E_\xi}= (k_F \xi)^2.
\end{equation}
Under the relation $k_F \xi \sim 1$, we can easily see that Fermi energy and the crossover energy to the universal low-temperature regime are about the same and of the order of $10K$.




\subsection{Why fermions, not TLSs, is a consistent description of SNS} \label{fermions-not-TLS}

In this section I argue why localized fermions, not TLSs, is a consistent framework for LEEs in SNSs.

Firstly, a fundamental theory must explain consistently \textit{not only LEEs but also the ground state}. AHVP theory comes short of such a requirement. Although AHVP model describes successfully the properties of LEEs, it does not describe the ground state of T-TLSs. 
In particular, AHVP is silent about why the density of TLSs is of particular value.

GPP model does describe both, the ground state and LEEs, in a single framework, even though not microscopically. 
GPP, similarly to AHVP, describes LEEs, which explain the specific heat and contribute to the scattering of phonons (Eqn. \ref{kappa-thermal}) and the internal friction coefficient (Eqn. \ref{Q-transverse}).   
Yet still as importantly, the ground state of SNSs is described by Fermi volume and Fermi density of glasson states, which enter into the calculations of the section (\ref{basic-consistency}).

The stability of the ground state of pseudo-spins proposed by AHVP can be questioned in the view of independent (i.e. bosonic) nature of different TLSs. The ensemble of anomalous LEEs is only a subsystem of the disordered network of atoms of SNSs, nevertheless this subsystem must be stable. The stability of bulk matter was accounted for by Pauli principle\cite{Lieb-review}. Since, in general, local LEEs can hybridize, get larger in size, and change their spatial positions, the stability of the ensemble of bosonic (and dense) LEEs is highly questionable. The historical works by Lieb\cite{Lieb-review} and Dyson and Lenard\cite{Dyson-stability} are strongly suggestive that local LEEs are fermi excitations if these excitations are not fully ``pinned" locally. 

Secondly, AHVP model simply uses the constant-in-energy density of states as a fitting parameter. GPP model has  the constant-in-energy density of states near the Fermi energy due to the Pauli exclusion principle, while the magnitude of the density of states is derived from the existence of a single crossover length scale.
AHVP model leaves unanswered the question why TLSs can be considered as completely isolated and have a particular and constant density of states.

In the discussion of a general many-particle system, it is important to consider not only the level repulsion 
but also different symmetries of the ``perturbative'' Hamiltonian terms, which hybridize various many-body energy levels of original Hamiltonian. The symmetries include Fermi-Pauli occupation constraints as well as the form of various couplings (see Eqn. \ref{H_int}). The emergence of composite fermions-glassons can mean that some fermionic degrees of freedom, plentiful at higher energies (for instance, electrons), survive under mixing effects of disorder and various effective interactions.

Thirdly, AHVP model attributes the energy relaxation to a local tunneling mechanism rather than collective multi-scale mechanism typical for all classical and quantum types of glasses. 
From the early days of theoretical analysis, it became clear that AVHP model cannot explain all experiments naturally with a single type of T-TLSs\cite{Black-Halperin}. 
In the framework of GPP, the calculations (Eqn's \ref{xi-m}, \ref{Q-transverse}, \ref{p_F}, \ref{m_F}) show that dipolar interactions and Fermi statistics are important. Fermi glass with dipolar interactions and Fermi statistics constraints appears to show collective relaxation mechanism\cite{Imry-2017}, which can be explored in the future work. Various glasses\cite{Imry-2012,  magFermiGlass}, including electron glasses, do show logarithmic relaxations independently of the nature of local LEEs, which can be electrons, spins, or TLSs. 
Fermi glass and TLS glass seem to belong to the different universality classes\cite{Imry-2017}, and this is how GPP and AHVP theories might get distinguished.

\subsection{Open questions and further experiments} \label{open-questions}

Some questions are beyond the initial investigation of this paper. In particular, a detailed microscopic understanding of glasson excitations is beyond the scope of the paper. 
GPP explains the universality of SNSs by the means of collective formation of composite fermions-glassons and their weak interactions with each other and with phonons. 
Fermi volume of glassons is small $(k_F a)^3 \ll 1$ and determined by MRO length $\xi \simeq k_F^{-1}$. 
The work, conceptually related to GPP, by Vural-Leggett\cite{AJL-V-RG} and more recently by Shukla\cite{Shukla} aims to derive the universal properties of SNSs microscopically. In particular, Shukla\cite{Shukla} relates the universality to the universal ratio of MRO $\xi$ to the `closest approach' inter-atomic distance $a$. The important role of this small ratio $a/\xi$ was highlighted in my previous work\cite{Turlakov} based on quasi-classical viscoelastic approach to the density-density correlation function. The current work here starts from fundamentally quantum perspective, yet the universality of internal friction $Q^{-1}$ relates to the same structural ratio of $a/\xi \simeq k_F a$. 

An open question is a detailed relation between a single crossover length $\xi$ as discussed in (\ref{length-scaling-crossover}) and medium-range order length as seen in experiments from the early days of the field\cite{Elliott}.
Spectroscopic measurements, related to this open question, should shed light on the connection between Boson peak modes ($E_{BP} \sim 4meV \simeq 40K$ for concreteness) and low-energy modes (T-TLS, glassons, etc.), which was pointed out by several authors\cite{Turlakov, Parshin-TLS-BPeak, Baggioli-2019}. 
Ioffe-Regel limit for transverse phonons is an important phenomenon for understanding the properties of SNSs. An important contribution to Ioffe-Regel condition should come from anharmonic coupling between transverse phonons and glassons when their momenta match $k_{ph} \sim k_F$. This latter consideration might reveal some relations between close energy scales $\epsilon_F$, $E_{int}$, and $E_{BP}$ to each other.

An important experiment using resonant sound absorption has already been proposed\cite{AJL-V} in order to test the isolated nature of just two energy levels of TLSs as long as local intra-site transitions are strictly dominant that inter-site transitions.
The linear resonant response of TLSs and localised fermions seems to be the same  under the same assumption. Therefore only the non-linear response as a function of sound intensity might distinguish between TLSs and localised fermions\cite{Glasson-experiments}. 

Promising experimental advances in low-temperature glasses in last 10 years are due to experiments on superconducting circuits encased by aluminium oxide layers, which are amorphous insulating solids. These experimental works aim to understand the microscopic nature of LEEs (or TLSs) directly\cite{Muller-circuits} as well as interactions between LEEs\cite{Lisenfeld-interacting, Burnett-extended}. The latter experiments address electric dipole moment of LLEs, while this paper considers only elastic dipole moment of LEEs, mainly due to the associated universal sound absorption by elastic dipole moments. However it is well-known that LEEs have both, elastic and electric, dipole moments. One open question is the strength ratio between elastic and electric dipole moments depending on chemical and physical structure of SNSs. Another open question is whether glassons, being fermions, might emerge from ``dressed electrons''(see, for instance, \cite{Agarwal-IvarMartin}). To this end, a glasson, being a composite fermion, acquires its fermionic nature from an electron. This composite heavy electron, i.e. \textit{glasson, is medium-range delocalized from local chemical bonds up to length scale $\xi$, yet it is strongly localized beyond the length $\xi$.}



Spectroscopic experiments in the context of superconducting circuits\cite{Muller-circuits} are promising local probes. These experiments\cite{Burnett-extended} can test whether the spatial size of local LEEs is of the order of crossover length $\xi$. These experiments\cite{Lisenfeld-interacting} should be able to distinguish between \textit{two interacting pseudo-spins (i.e. TLSs) and two interacting fermions,} since the effective form of fermion-fermion interaction is quite different from spin-spin interaction\cite{Glasson-experiments}. It will be necessary to distinguish short-range and long-range (dipole) parts of the interaction as well as elastic-dipole and electric-dipole parts.
But the advantage of such experiments is that they measure individual interaction rather than statistically averaged effects of many TLSs.

Non-interacting Fermi glass has a well-defined locus of the Fermi energy in the momentum space. It remains to be tested and examined whether weakly-interacting \textit{strong-disorder Fermi glass}, as conjectured for SNSs here, still has a detectable locus of the Fermi surface. Nevertheless some spectroscopic features of the Fermi energy would provide a direct evidence for the existence of glassons, because such a finding is indeed surprising from the point of view of the Standard Model\cite{anderson, Phillips}.

\section{Conclusion} \label{conclusion}

On the basis of fundamental spin-statistics theorem, I propose that \textit{infamous LEEs in SNSs are localized fermions.} These fermions have intrinsic half-integer spin as a quantum number, and thus they do not hybridize directly with phonons. Fermions-glassons are stress-field excitations of coarse-grained isotropic blobs with the size equal or larger than MRO length $\xi$.
Localized and weakly interacting glassons form a Fermi glass in the ground state. Glassons can be loosely thought as an alternative version of TLSs respecting quantum statistics, yet with important, experimentally distinguishable, differences (see \ref{fermions-not-TLS}).

The fermion phenomenological theory proposes fermionic LEEs and the ground state of Fermi glass as an explanation of low-temperature SNSs. 
Glasson parameters are calculated theoretically (Eqn. \ref{xi-m}) and from experimental data (Eqns. \ref{p_F} and \ref{m_F}) and found consistent.
The glasson-phonon phenomenology (GPP) explains (Eqns. \ref{kappa-thermal} and \ref{Q-transverse}) the experimental specific heat due to glassons, phonon thermal conductivity limited by glasson-phonon scattering as well as the small internal sound friction coefficient $Q^{-1} \sim (k_F a)^3$.   
The constant-in-energy density of states near the Fermi energy is a natural consequence of the fermion phenomenology unlike being a free-fitting parameter for AHVP model.


In the part (\ref{Fermi-parameters-experiments}), I present a basic calculation of Fermi volume and mass of glassons directly from the linear-temperature slope of the specific heat and conjectured Fermi energy. This calculation, which can be considered on its own, shows simple consistency (Eqn. \ref{p_F}) between plausible value of Fermi energy and small Fermi volume. 

The crossover (as well as medium-range-order) length $\xi$ determines the small wavevector (radius) $k_F$ of glasson Fermi volume due to the condition $k_F \xi \simeq 1$. 
\textit{Weak glasson-glasson and glasson-phonon interaction constants are also the consequence of the small parameter $a/\xi \ll 1$.} 
The glasson mass $m_{gl}$ is found to be somewhere between electron $m_e$ and atom $M_{atom}$ masses, $m_e < m_{gl} < M_{atom} $. 


“Everything should be made as simple as possible, but no simpler.” By proposing LEEs to be fermions I rely on Occam's razor principle.
The simplicity of the fermion phenomenology holds promise of tangible progress beyond the picture of generic collective modes\cite{Yu-Leggett,AJL-V-RG,Turlakov}.

\textbf{Acknowledgements} 

I am grateful for helpful discussions to Ivar Martin (Argonne National Lab) and Peter B. Littlewood (University of Chicago). 


\bibliography{glassons}

\bibliographystyle{unsrtnat}

\end{large}
\end{document}